Astro2010 Technology White Paper

# Large Focal Plane Arrays for Future Missions

Paul Scowen[+](paul.scowen@asu.edu) & Shouleh Nikzad *(shouleh.nikzad@jpl.nasa.gov)

[+]School of Earth and Space Exploration, Arizona State University
Tempe, AZ 85287-1404
Tel:(480) 965-0938

and

Michael Hoenk, Ivair Gontijo, Andrew Shapiro, Frank Greer, Todd Jones, Suresh Seshadri, Blake Jacquot, Steve Monacos, Doug Lisman, Matthew Dickie, and Jordana Blacksberg

*Jet Propulsion Laboratory, California Institute of Technology
Pasadena, CA, 91109

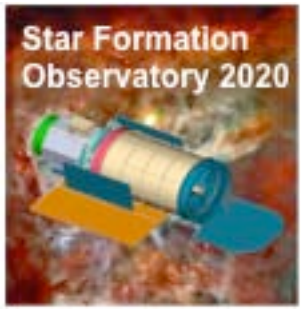

SFC science

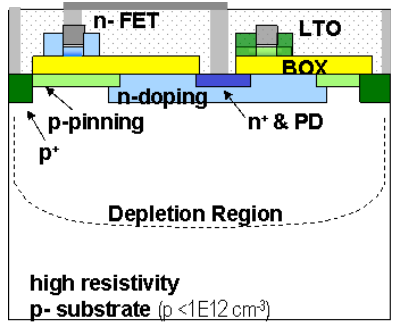

CMOS-SOI Imager

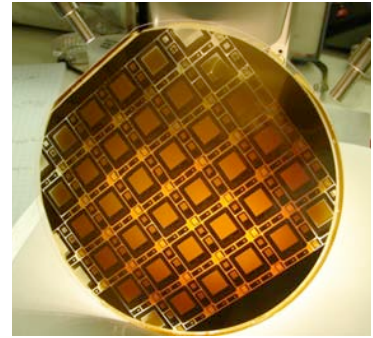

Precision back-thinning of CCDs & CMOS focal planes

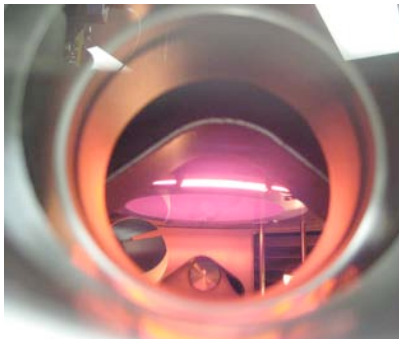

High-throughput, wafer level delta-doping of CCD & CMOS focal planes

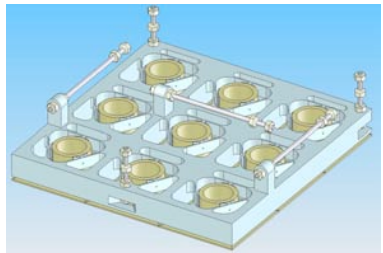

Mosaic Focal Plane Array partially filled with detector raft assemblies

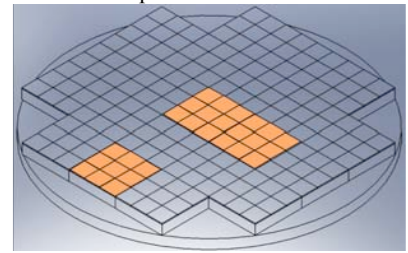

A 3x3 raft of modular detector assemblies

**Abstract**

We outline the challenges associated with the development and construction of large focal plane arrays for use both on the ground and in space. Using lessons learned from existing JPL-led and ASU/JPL partnership efforts to develop technology for, and design such arrays and imagers for large focal planes, we enumerate here the remaining problems that need to be solved to make such a venture viable. Technologies we consider vital for further development include: (1) architectures, processes, circuits, and readout solutions for production and integration of four-side buttable, low-cost, high-fidelity, high-performance, and high-reliability CCD and CMOS imagers; (2) modular, four-side buttable packaging of CCD/CMOS imagers; (3) techniques and hardware to test and characterize the large number of chips required to produce the hundreds of flight-grade detectors needed for large focal-plane missions being conceived at this time; (4) ground based testbed needs, such as a large format camera mounted on a ground-based telescope, to field test the detectors and the focal plane technology solutions; and (5) validation of critical sub-components of the design on a balloon mission to ensure their flight-readiness. This paper outlines the steps required to provide a mature solution to the astronomical community with a minimal investment, building on years of planning and investments already completed at JPL.

**Introduction**

As we move into the next decade in astronomy, it is becoming very apparent that there are a large variety of scientific applications that would benefit dramatically from access to truly large focal plane arrays combined with very small pixel plate scales to provide high angular resolution in tandem to the large areal sample. This paper discusses a few exemplar scientific drivers for such a capability as well as lays out the technological challenges that have already been studied as part of existing missions and future mission concept studies. We believe this is an area of technological development that is primed for truly innovative advancement with a relatively modest investment over the next 10 years, and the benefit to the scientific community would be considerable.

To support sweeping new science programs, we need to develop the technological capability to construct large focal plane arrays (FPAs) that are flight rated for space in a reliable and straightforward fashion that simultaneously mitigates risk, maximizes yield, and minimizes cost. The scope of this challenge is significant in that it will likely require many years of concerted effort and investment by the entire astronomy community to develop new FPA design, fabrication, and assembly technologies; data handling and storage methodologies; and observational strategies to allow a large scale systems to be built in a routine fashion.

In this paper we outline the specific challenges associated with development of large FPAs for future missions and lay out a possible path forward along with cost projections to achieve those goals.

A. *The Case for the development of large FPAs:* Specific science drivers

While large FPAs have general appeal in future survey missions, we use a representative example to define the science drivers. The *Star Formation Camera* (*SFC*) is a concept design for a wide field high resolution UV/optical camera to be flown in space on a large aperture (4m) telescope to execute survey science that was studied as part of NASA's ASMCS input to the Decadal Survey.

We believe there is a compelling case for a comprehensive UV/optical/NIR wide field imaging survey of Galactic star formation regions that can probe all aspects of the star formation process. The primary goal of such a study is to understand the evolution of circumstellar protoplanetary disks and other detailed aspects of star formation in a wide variety of different environments.

| Table 1: Baseline SFC focal plane | | |
|---|---|---|
| Quantity | Units | Original (revised) specs |
| SFC focal plane | # of arrays | 540 |
| Array format | # pixels | 3.5k x 3.5k |
| # of parallel outputs | | 4 |
| Full well | Ke- | 130 |
| Read noise | e- | 3 |
| Dark current | e-/pix/hr | 0.5-2.0 (10) |
| Operating temperature | $^o$K | 133 (175) |
| Readout rate | kHz | 50/output |
| Readout time | seconds | 61 |

This requires a comprehensive emission-line survey of nearby star-forming regions in the Milky Way, where a high spatial resolution telescope+camera will be capable of resolving circumstellar material and shock structures. Such data will allow production of precise color-color and color magnitude diagrams for millions of young stars, and for the first time we could systematically map the detailed excitation structure of HII regions, stellar winds, supernova remnants, and supershells/superbubbles. This survey will provide the basic data required to understand star formation as a fundamental astrophysical process controlling the evolution of the baryonic contents of the Universe. The majority of the tracers being targeted require the angular resolution and wavelength agility of a medium to large aperture (1.5-4m) UV/optical space telescope combined with a wide-field imaging camera. The *SFC* design and observing program assumed the use of CCD arrays with certain assumptions and requirements on its performance presented in **Table 1**.

B.  *Suitable architecture for large FPAs*

The design of large focal plane arrays is driven by several factors. One is the set of scientifically-driven requirements for the detectors themselves. Such requirements can include the required passbands for observation, the form factor and pitch of the detectors, the overall size of the array or field of view of the camera, the seam size, the detector clocking rate, the noise characteristics and therefore the operational temperature of the detectors, and considerations for total data storage, compression and transmission. Another factor is that of the electronic architecture that needs to be adopted to minimize the risk associated with integration and testing of the focal plane as it is constructed and prepared for flight.

Development of modular designs for the sensor, its control and readout electronics are recommended, to allow individual detectors to be tested and easily removed as needed, once integrated into the focal plane assembly, with a minimum of impact on any neighboring detectors. Each channel needs to be designed with a dual string layout to allow for failure during the mission, and each chip needs to have a power supply that is as noise free as possible.

By comparing the needs of real world large FPA missions, ranging from the *SFC* concept study to the recently launched *Kepler* mission, several system architecture issues arise. The 95 Megapixel *Kepler* Focal Plane Array Assembly (FPAA) is populated with 46 CCD-based detectors of which four provide fine guidance control. The 42 science CCDs are arranged as 21 CCD modules, with two CCDs housed per module. An electronic rack of boards lies immediately behind the *Kepler* FPA and supplies bias and clock inputs and pre-amplification of CCD output signals. The *Kepler* FPA is cooled using dual constant conductance heat pipes and a dedicated passive radiator. The *Kepler* system power is 130W, of which 7W is for the FPA power. *Extrapolating this number for our CCD-based SFC observatory design places a requirement of ~35 W of dissipation directly on the focal plane and 650 W for the entire system if we used the Kepler paradigm. The SFC solution also uses variable conductance heat pipes (VCHPs) have been baselined to transport heat from each FPA to the cryogenic radiator – the two SFC FPAs are baffled from thermal radiation from the surrounding optical instrument by a two-level shield since it was found the dominant heat source was radiative loading from the room temperature instrument enclosure.* Additionally, the 57 Kg mass of the *Kepler* substrate/electronic card cage assembly and wire density places additional burdens on the mechanical rigidity of the FPA. Such a design methodology becomes unwieldy for *SFC*-scale tiled FPA arrays that use many more imagers in their FPAs. *An alternative approach, considering replacement of the electronic rack with Application Specific Integrated Circuit (ASIC) controllers such as has been done for the JDEM/SNAP project ameliorates the power/mass/volume/cost for CCD focal planes, but does not constitute a globally optimal solution as a fully integrated, low power focal plane, as might be achieved with CMOS imagers.*

C.  *Focal plane scaling issues*

The overall size of the FPA drives a series of physical and programmatic issues:

- Power.  What level of power is required by the FPA during integration and readout in a reasonable period of time?
- Thermal control of large numbers of detectors in a space-based environment – what is the right balance between thermal demands for power and heat management when compared with the effect on detector performance in terms of acceptable dark noise?
- Yield for production of large numbers of flight-rated detectors.  If this yield is sufficiently low, what is the break-even point between cost of purchasing a large number of lot runs versus the relaxation of specifications for CCD performance in terms of cosmetics and DQE stability and uniformity across the focal plane?
- Risk.  What level of risk is acceptable for arrays of literally hundreds of arrays - what new paradigm must we adopt when moving into the era of hardware development and construction of large FPAs?
- Data Volume.  What capacity of data volume can we safely store and read down from orbit to the ground with minimal loss, maximal fidelity, and with sufficient speed not to undermine any survey goals – large focal planes will be used primarily for survey science?
- Areal Coverage and Completeness.  What can be done about inter-chip gaps?  Can this be minimized over such large areas?  What level of flatness can be maintained for truly large focal planes – what does this do to specifications about depth of focus for the optical system and the tolerances on manufacture and assembly of the focal plane itself?

D.  *Control of observational efficiency and flow-down detector requirements*

The grid on the left of **Figure 1** schematically illustrates an *SFC* focal plane. The white lines making up the grid represent gaps between individual elements of the imager mosaic. Such gaps, which in today's imagers can be as large as 100s of pixels, are required for mechanical tolerance during wafer dicing and focal plane assembly, field shaping to control dark current, presence of column or row circuitry, etc. Any core science observing program will, therefore, need to include some level of dithering to fill in these gaps and to be able to reject and correct defects that are stationary in the detector coordinate frame (e.g., hot and/or unresponsive pixels, less responsive pixels, multi-pixel blemishes, etc.). The right side of **Figure 1** illustrates a dithering strategy (over distances comparable with the size of these seams) that recovers the full areal coverage and corrects for stationary detector defects. The survey tile is produced by co-adding the FPA images at the 3 'major' pattern pointings within an observing block, and then is shifted back to a common pointing. However, as shown in the figure, the impact of such a strategy on the efficiency and cost to the survey schedule is large. *An area of technological development that would pay high dividends over the next decade would be the minimization of inter-chip seams which are a common burden on the scientific use of large focal plane arrays.*

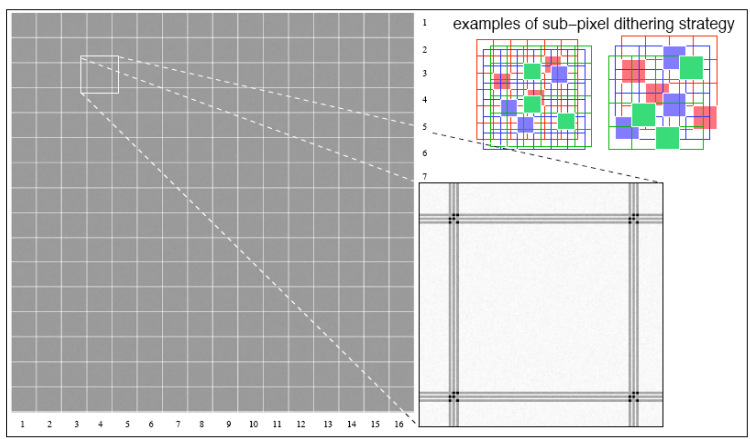

**Figure 1**. The *SFC* FPA mosaic or stack of nine individual exposures at 3 different pattern pointings (with three subpixel dither/CRSPLIT exposures) that is our assumed baseline survey strategy to provide full areal coverage, reject detector defects and cosmic rays, and reconstruct the PSF delivered by the 4 m *Theia* OTA.  With this strategy, 92.2% of the area will have the full exposure time, 7.7% of the area will have 2/3 of the full exposure time, and only 0.1% will have 1/3 of that time.

Limitations of the imager technology itself can also significantly impact the efficiency of the *SFC* science observation, alternatively defined as the fraction of the total elapsed calendar time that the shutter was open. For example, although for the survey it was assumed each CCD would be clocked out in parallel at 50 kHz through 4 amplifiers per CCD for a total read-out time of 61 sec, the hardware could support clocking rates up to 100 kHz. A 32 sec read-out time has distinct advantages in terms of survey duration and would raise the survey efficiency from 65% to 74%. Increasing the read-out speed to 100 kHz would require roughly twice the power needed for reading at 50 kHz (which does not take the warm electronics into account; only the CCDs themselves). A critical consideration, however, would be whether or not the read-out noise would indeed still remain at or below 3.0 e−. Preliminary indications are that the read-noise at 100 kHz read-out might be 3.7 e− which would raise the mean exposure times necessary through the planned filters from 86 and 252 s to 98 and 293 s, typically. As such the additional exposure time required for this change nearly cancels the benefits achieved from the shorter read-out time. *This trade-off can be improved in 5–10 years time with the development of CCDs and other imaging technologies with better noise bandwidth on a per pixel basis.*

E.  *Populating the focal plane: Detector options*

Two major classes of detectors exist in the UV/optical/NIR spectral range: substrate-removed HgCdTe and silicon imagers. While high quality, large format HgCdTe FPAs are commercially available, the high cost of these FPAs remains a major disincentive to their use in large focal planes that do not require spectral response beyond 1 µm. The maturity of silicon imagers and their high performance and recent development to extend the spectral range in both long and short wavelength renders silicon imagers as detector of choice for UV/optical/NIR instruments.

Silicon imagers can be divided into those with a high degree of integration (e.g. CMOS imaging arrays with on-chip bias, timing, control and many parallel readouts), CCDs (serial readout, charge transfer to final one or few amplifiers), and hybrid arrays (parallel readout, PIN diode arrays bump bonded to a CMOS readout). Additional new monolithic approaches including CMOS-SOI imagers are being developed at JPL and other institutions to address the need for separation of the detector and readout optimization while removing the cost and reliability concerns associated with hybridized approach using bump bonding. Such imager technologies offer the potential for high degree of integration to reduce the system mass and cost, low-power to reduce the thermal load, high-yield to lower procurement cost and high-reliability to preserve mission life. Choice of the detector architecture will largely depend on the science requirements and instrument design. While existing CCD options likely have sufficient performance, CTE, power dissipation, integration and cost are primary reasons for the development of alternative imaging technologies such as CMOS. The CMOS FPAs will have to have performance comparable to CCDs. Areas of development include: i) devices offering wide spectral bandwidth (long wavelength response to λ ≥ 1 µm and ii) photon counting (devices with sub-electron noise and/or photoconductive gain).

Regardless of readout schemes, to achieve the highest performance in quantum efficiency (QE), spectral range, fill factor, and dark current (surface-generated), silicon imagers must be back illuminated with proper processes such as JPL-invented delta doping technology. In moving towards the goal of routinely fabricating large numbers of chips for such focal plane arrays, JPL has recently invested in equipment and infrastructure for high throughput, high yield production of back illuminated, delta-doped imagers. This versatile capability enables back-illuminated detector production for large FPAs such as *SFC* within mission cost and schedule. Additionally, JPL has invested in the design of novel detectors and packaging of detector modules and focal planes in anticipation of future missions requiring high performance large format FPAs.

F. Packaging and assembly

A conceptual mosaic packaging assembly is depicted in **Figures 2 & 3**. The modular construction permits any single detector to be easily and independently removed and replaced without adversely affecting other detectors in the array. This modularity of the tiles in a large mosaic focal plane and the ability to replace or troubleshoot individual modules will affect the cost and reliability of the instrument. CCD, CMOS imagers, and hybrid arrays that could populate the FPA, can employ bump bonding between an imager in the array and its associated control/readout electronics for tight integration. Alternatively, flex cabling for this interface can be used to provide separation and minimize the electronics thermal load on the FPA.

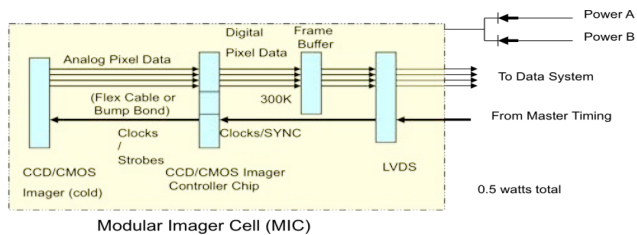

**Figure 2**. An example of a detector modular unit for a large focal plane array. Such a design allows for simple swap-out of failed detectors, without impact to adjacent detectors and the backplane electronics.

Array tile alignment and systems packaging is critical for successful large focal plane arrays. The tile seam width directly affects science return and observational strategies. Completed detectors also have to be mated with a mechanical interface that enables the detector to meet the optical specifications of the instrument. Typical FPA planarity requirements at present are on the order of ±10 µm and the XY location of each detector will need to be controlled to a few tens of microns.

Mounting of the detector onto a stable interface is expected to accommodate and remove any departures from planarity inherent in free-standing processed silicon wafers. The mounted detector array has to meet the optical specifications at its operating temperature and it has to survive ground testing and orbital temperature cycling extending over the life of the mission. We believe that while metrology will be needed to assure that opto-mechanical specs have been met for the fully assembled focal plane, the mechanical tolerances required to meet the optical specifications are better addressed within the individual modules, during their design and fabrication, rather than at the focal plane during assembly.

Some of the above challenges have been faced in previous FPAs such as *Kepler* but there remains much room for improvement in areas such as power and tile seam size. In other missions such as *Herschel-Planck* and the *Mars Science Laboratory*, array packaging has been developed by JPL using different paradigms. Techniques involving sub-micron optical alignment, laser welding and hermetic packages have been developed and demonstrated over many (>2000) extreme temperature (down to mK to 500°C) cycles in these and developmental endeavors.

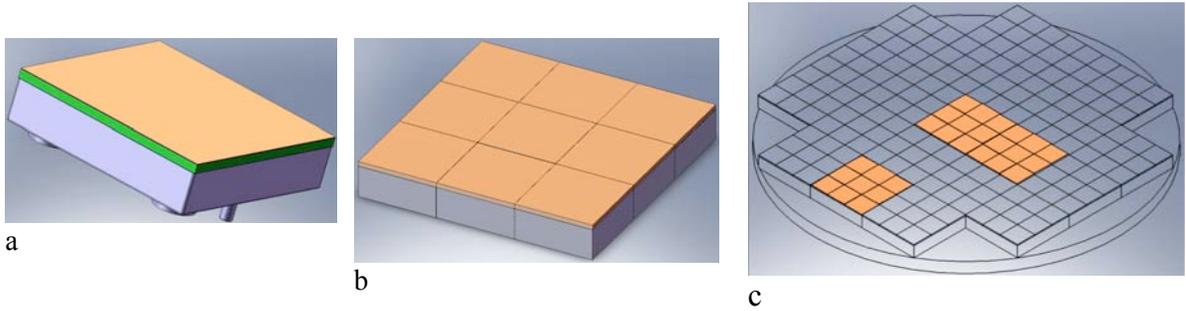

a
b
c

**Figure 3**. Schematic of large FPA: From left to right, Individual detector module, Raft of 3x3 detector modules, and full FPA

G.    *Controlling the data volume*

With most detectors now requiring 16-bit readout to generate enough dynamic range for most scientific problems, it is easy to see that where daily data volumes are going to approach several terabytes within the next decade. This means that another area of technological development that needs investment is the provision of large volume, solid-state storage that is flight rated, the continued development of faster computational units for orbit, and the enlargement of data transmission technology from orbital locations as far away as L2 (which is preferred for a series of mission planning and maintenance reasons). *Toward mitigating future memory requirements, JPL has been developing a concept called SyFT/PRODIGY to implement a smart tiled FPA that can identify, zoom in and track target events of interest at the full resolution of the FPA in real-time based on search parameters programmed into the system. Such an approach will not only provide for intelligent data mining but also better science by training the "attention" of the instrument to the targets of interest.*

H.    *Impact of Detector Fabrication Process Yield and Acceptance Criteria*

In an era of large FPAs, the required number of flight-rated detectors will be substantially larger than was needed for current missions. Using the *SFC* instrument concept as an example, 540 flight-rated detectors will be needed to successfully tile the two required focal planes. This is an order of magnitude larger than the *Kepler* mission. Complicating matters is the set of strict cosmetic and performance-based selection criteria that are commonly used today to select detectors for flight. It has been typical therefore to see many lot runs purchased to yield as few as 4 flight-rated detectors, resulting in a net yield of <10%. Clearly, without investment in new detector fabrication methodologies and infrastructure, these twin constraints of large detector quantities and low yields would result in extremely high costs and prohibitively long delivery schedules. To address this issue, JPL has recently embarked on a sizeable effort to scale up its UV/Vis/NIR back-illuminated detector fabrication processes and facility infrastructure to meet these demanding large FPA mission deliverables. This infrastructure includes JPL's acquisition of a state-of-the-art Si-MBE system that enables up to 8-inch wafer level processing of detectors, as well as batch processing of multiple wafers to the point where a lot run of wafers can be processed in record time. This capability *dramatically increases the throughput and reproducibility over the equivalent die level processes.* As 10's or 100's of detectors can be processed at the same time, wafer level processing alone results in almost an order of magnitude reduction in the time and labor costs required to fabricate scientific grade detectors. Wafer level processing also reduces variability between detectors because they are processed together. These tighter distributions in device quality enable one to systematically improve yield through process optimization. Thus, more high quality detectors are produced in each run, and the time to complete a run is reduced. In parallel with these process and infrastructure development investments, the impact of less restrictive flight device selection criteria should also be explored. For example, relaxing the selection criteria to include devices with more bad columns, hot pixels, and pixel-to-pixel variations in DQE that depart from typical specs of 3% or less would clearly reduce mission costs. Such device imperfections could be dealt with by an expanded and more creative calibration program to minimize the risk to mission science goals. For example, a foveal construct can be employed where "perfect" devices would populate the most important central position(s) in the FPA. The lesser performing devices may then be kept for fringe locations. This could improve the yield and thereby reduce the cost while improving the observational capability of the instrument along with SyFT/PRODIGY paradigm. The large FPA era requires new paradigms for device fabrication and selection, and forward-thinking criteria to produce and down select devices for flight while keeping the mission costs and risk minimal.

## I.  Testing

While the construction of large numbers of detectors for large focal plane arrays is a challenging problem, the subsequent testing of those same numbers of detectors can potentially be overwhelming. The definition of a straightforward, quick and efficient testing program is critical to making the construction of such arrays even plausible. One can envision three levels of testing of individual focal planes: (1) detailed characterization of a small number of devices to determine optimize device operation and evaluate performance trade-offs; (2) screening of a very large number of devices; and (3) characterization of screened devices to bin them for flight, engineering, spares and rejects.

The development of automated cryogenic test systems (ACTS) will be essential for both quality control and throughput for these tests. In addition to addressing a list of detector performance issues, such test systems must also be configured to address new issues such as inter-operability between the detector and its control electronics and between these detector electronics and those off the focal plane. Such systems must also be capable of operating multiple detectors side by side to demonstrate that interference (crosstalk) is acceptable.

Experience strongly suggests the need to develop experimental testbeds to exercise the focal planes under simulated observation conditions to verify operational protocols, mitigate or calibrate undesirable characteristics and determine relevant selection criteria for engineering and flight qualified devices. The suggested effort is a new capability that is traditionally NOT part of detector performance testing.

*Large Focal Plane Arrays for Future Missions*J.     *Flight readiness of technology: How to raise TRL*

It is the goal of any development program to raise the TRL of all components and systems to TRL 6 to mitigate the risk associated with adoption for flight. Therefore, in addition to working on process improvements and innovations for meeting FPA requirements at low cost, low power, and low mass, it is imperative to work with the community toward suborbital flights and ground observations.

The JPL Gigapixel Initiative is currently moving toward high capacity manufacture and processing/doping of wafers and sets of CCD or CMOS chips. *A logical and necessary step between these pieces and finally producing detectors of high TRL is the use of such chips in the field.* We believe that a necessary step is to develop a wide-field ground-based camera to get real and practical data about the challenges of making large focal plane arrays work together when making real astronomical observations. Questions such a demonstration would answer include: can we read out all the chips without crosstalk, with stable enough amplifiers and ADCs to generate science-grade data? Can we manage the thermal stability of the system? Can we handle the data loads generated? Beyond that we also need to demonstrate the durability and dependability of the detectors in flight situations. *We contend that a vital piece of such a plan is to fly detectors on balloon missions to observe their performance in near-space environments to answer questions such as:* How is the thermal problem different in space – what does that do to the electronics, the mechanical rigidity of the system, the performance of the chips themselves and the durability of the technology to physical abuse from launch to landing? We believe these steps are vital to extend the development process beyond what is possible in the lab on an optical bench and to move towards reliable production of large FPAs of *SFC* scale (100's of detectors). This kind of assembly has the potential to become the production line workhorse for all the scientific uses the community has in mind over the next 10-20 years.

K.     *Cost and Schedule*

The cost to develop and demonstrate a sufficiently large format unit cell of a new detector technology requires several million dollars. We have developed an outline for what we believe is necessary to move from the current state of the art to a production line environment for twin mosaic FPAs with as many as 540 separate flight-rated detectors, as would be needed by *SFC*. Shown in **Table 2** are the rough order of magnitude (ROM) cost and schedule of development and TRL advancement of this large FPA. We have used some of the cost estimates obtained in the Team X study for *SFC* as well as figures based on experience and information at hand today.

| Table 2. ROM Cost and schedule for technology development and TRL advancement. | | |
|---|---|---|
| Tasks | Cost | Duration |
| Complete infrastructure for processes and facilities | $5M | One year |
| Fabrication and validation of prototype detector modules | $10M | Two years |
| Fabrication & validation and demonstration at a ground observatory of a prototype 3x3 raft modules | $10M | Two years |
| Fabrication and validation of prototype FPA | $10M | One year |
| Detector Balloon Demonstration (parallel w/ steps 1-3) | $5M | Four years (1$^{st}$ flight) |

**Grand ROM Total: $40M**

*Scowen, Nikzad et al.*                                                                                     9

Clearly, this level of development requires, at the very least, a supplemental provision within the current ROSES APRA framework because of its cost cap. The alternative is to fold in such development at a very low TRL into a mission schedule, thereby leading to increased mission, schedule and cost risk.

## Recommendations

We have reviewed the scientific drivers for large FPAs and the technological challenges associated with them. As a community we are approaching the need to access the unique parameter space provided by the combination of both wide field imagers (tens of arcminutes on a side) with the highest resolution we can reasonably achieve (10's of mas) to be able to study statistically complete samples of objects whose variability and population is unmeasured, and to gain access to sufficiently rare objects of small enough size that only with this combination can we ever hope to find enough examples to understand them. To achieve such capabilities we have to surmount technological challenges. Many of these require investment over the next decade to make this dream a reality. The issues that need particular attention are:

- Assessment studies to develop point designs that cover at least the extremes of likely science applications (e.g. true imaging vs *GAIA*-like photometry) to determine commonalities of architectures needed for various missions and examine trade-offs with regard to the proper FPA architecture, size, mass, power and thermal control in space-based environments
- Development of high-throughput, high-yield back-illuminated delta-doping fabrication technology for low dark current and increased spectral response
- Development of high yield production of detectors, new "foveal" paradigms in populating FPAs, critical assessment of what changes in acceptable specifications for flight-rated detectors in large FPAs as compared with the era of single-detector selection for flight missions
- Development of scalable, low-cost, large format imager technologies and arrays that meet the focal plane architecture requirements while optimizing power, speed, noise, resolution, spectral response, imager gap, depletion range, and fill factor of the imager chip
- Development of modular packaging architectures and fabrication techniques that minimize seam size between imager modules, meet the mechanical and optical tolerances and metrology requirements of the fully assembled focal plane and enable easy placement/test/removal/replacement of modules during FPA integration
- Development of high capacity testing capability to reduce the lead time to construct such arrays, while preserving the fidelity of the product and mitigating risk associated with fabricating the final array
- Development of high-capacity data storage, compression and transmission technology to accommodate terabytes of data such FPAs are likely to produce on very short timescales
- Investment in opportunities to provide a path to raise the TRL using ground based telescopes and suborbital programs to produce a low-risk solution to the problem of detector selection and system evaluation